\documentclass{kluwer}
\input{psfig}				%

\def\saxjdhdn{\mbox{SAX J1819.3-2525}}
\def\asec{^{\prime \prime}}
\def\kms{\mbox{ km\,s}^{-1}}
\def\nh{N \mbox{(H)}}
\def\cmmoinsdeux{\mbox{ cm}^{-2}}
\def\kpc{\mbox{ kpc}}
\def\Msol{\mbox{ }M_{\odot}}

\begin{document}
\begin{article}
\begin{opening}
\title{On the nature of the microquasar \texttt{V4641 Sagittarii}\thanks{Based on observations collected at the European Southern Observatory, Chile. 
S.C. thanks Rob Hynes for having alerted him to this new
flaring source on September 1999, and Bob Hjellming
for the communications he gave on the radio
observations.
S.C. is very grateful to the ESO/NTT team staff 
for their availability and skills.
S.C. acknowledges
support from grant F/00-180/A from the Leverhulme Trust.
IFM acknowledges partial 
support from Conicet/Argentina.}}

\author{S. \surname{Chaty}\email{s.chaty@open.ac.uk}}
\institute{Department of Physics and Astronomy, The Open University, 
 United Kingdom}
\author{I.F. \surname{Mirabel}} 
\institute{Service d'Astrophysique, CEA/Saclay, 
France \& IAFE/Conicet, Argentina}
\author{J. \surname{Mart\'{\i}}}
\institute{Departamento de F\'{\i}sica, Escuela Polit\'ecnica Superior, 
	Universidad de Ja\'en, Spain}
\author{L.F. \surname{Rodr\'{\i}guez}}
\institute{Instituto de Astronom\'{\i}a, Campus UNAM, Morelia, 
	Michoac\'an 58190, M\'exico}
\institute{}

\runningtitle{On the nature of the microquasar V4641 Sagittarii}
\runningauthor{Chaty S.}

\begin{ao}
Department of Physics and Astronomy, The Open University, 
Walton Hall, MK7 6AA, Milton Keynes, 
United Kingdom
\end{ao}

\begin{abstract} 
We present photometric and spectroscopic optical and near-infrared (NIR)
observations 
taken during the outburst on September
1999 of the source 
V4641 Sgr = SAX J1819.3-2525 \cite{in'tzand:2000} = XTE J1819-254 
\cite{markwardt:1999a}.
We show that our observations suggest a distance
 between 4 and 8 kpc, the spectral
type of the companion star being constrained between B3 and A2 V,
making the system a High Mass X-ray Binary System (HMXB).
In view of the radio and optical/NIR observations, it is possible that 
the ejecta of the source interacted with the surrounding medium
of the source.
\end{abstract}


\abbreviations{\abbrev{NIR}{Near-infrared};
   \abbrev{HMXB}{High Mass X-ray binary}}

\classification{JEL codes}{D24, L60, 047}
\end{opening}

\section{Introduction}

The source V4641 Sgr = $\saxjdhdn$ = XTE J1819-254 
attracted considerable attention after 
the detection of a giant optical outburst on 1999 September 15.7 UT, 
from the magnitudes 14.0 to 8.8 in the V-band \cite{stubbings:1999}.
The X-ray source flared, from 1.6 to 12.2 Crab in the 2-12 keV X-rays
on 1999, September 14th, as observed by XTE, 
through a brief but dramatic eruption, 
its position being coincident with
the optical transient \cite{smith:1999}.
Less than 10 hours later, the source was fainter than 50 mCrab.

The VLA detected on Sep. 16.0 UT
a strong radio source at 0.4 Jy at the
position of the variable star.
An elongation was
extending $0.25 \asec$ between 0.6-1.2 day after the X-ray flare,
and at the same position on 17.9, 22 and 24 UT.
This allows to classify the source as a new microquasar \cite{hjellming:2000}.
A detailed study of this source and details on the observations
are reported in \inlinecite{chaty:2001a}.



\vspace*{-0.5cm}

\section{Observations and results} 

The optical observations were performed 
with the NTT telescope and the instrument EMMI RILD.
We imaged the source in V, R, I and Z filters, and took
some spectra with the grism \#1.
The infrared observations 
were performed with the NTT and the instrument SOFI
through the filters J, H and Ks,
with an exposure time of 15 min. 
The optical and infrared lightcurves 
and the J-Ks color during the outburst
are reported in Figure 1.
In the optical, after the big outburst 
there was still some flaring activity through 0.5 mag in V, R, I with
no significant change in the colors.
There was also 
some flaring activity in NIR through 1 mag in J and K,
with a significant change in the J-Ks color during the post-outburst stage
(between 2 and 5 days after the giant burst).
This suggests an increased K-contribution compared to J,
either due to the emission of a jet, the appearance of heated dust, 
or even by the interaction 
with the interstellar medium.

We report the normalized
optical spectra offset in intensity to get an easier reading 
in Figure 2.
On a timescale of one day,
the lines were changing from emission to absorption.
All the Balmer serie is visible: H$\alpha$, $\beta$, $\gamma$, $\delta$, 
$\epsilon$, $\zeta$...
The H$\alpha$ emission line is extraordinarily strong: 
one day after the outburst, its
equivalent width was $\sim$ $100 \AA$, with a FWZI of 
$\sim 6700 \kms$ and a blue wing.
There was also a strong He I $5876 \AA$.
The Na-D gives E(B-V) = 0.25 implying 
$\nh = 0.13 \times 10^{22} \cmmoinsdeux$.
The strong variability of the lines 
and the blue continuum
suggest the emission from an accretion disk, or of a corona,
with accretion of matter onto a compact
object with a high-velocity wind component ($\sim 6000 \kms$), 
also perhaps the presence of a cocoon or a jet.


\vspace*{-0.5cm}

\section{Discussion} 


Taking the magnitudes at the faintest stage of the source 
in the optical and NIR wavelengths,
we plotted them on a color-magnitude diagram (Figure 3).
We also plotted three different absorptions, corresponding
respectively to $0.05$, $0.1$ and $0.15 \times 10^{22} \cmmoinsdeux$,
and different distances of the source,
from 1 to $10 \kpc$. We can see that 
the distance is constrained to
$4 < d < 8 \kpc$ in order to have a spectral stellar type
consistent with the stars of our Galaxy.
This location on the color-magnitude diagram suggests that
the spectral type of the companion star of the binary system is only 
consistent with an early type main sequence star B3 - A2 V.
Therefore the mass is constrained between 2 $<$ M $<$ 10 $\Msol$,
suggesting that it is a high-mass X-ray binary.
This is consistent with optical spectra in quiescence, 
taken by \inlinecite{orosz:2000}, who derived a stellar
spectral type of A2 V at a distance of $6.1 \kpc$.

If the elongation seen in the radio was a moving component, 
the proper motion is between $224 < \mu < 788$ mas/d depending on the
exact time of the ejection.
In the following we will assume that this is the
approaching (brighter) condensation with $\mu_a = 500$ mas/d.
Since $D\leq \frac{c}{\sqrt{\mu_a \mu_r}}$ 
and from our results $D \geq 4 \kpc$, we conclude that the apparent
velocity in the plane of the sky would be strongly superluminal,
$v_a$ of the order of 12.
However, 
no movement of this elongation was detected between Sept. 16.02 and 24.1 UT,
suggesting an interaction with surroundings at 
$0.25 \asec$ $\geq 1.5 \times 10^3$ AU at the distance of 6 kpc.
This is possible if the ejections began to take place 10 days before
the radio detection e.g. on September, $8^{th}$, and we can
see from Figure 1 that the source was already active
at this date.
It seems therefore that the activity of this source was not as sporadic
as we could have thought at the beginning.
Indeed, {\it RXTE} could detect this source during 270 days
before the outburst (in't Zand; Markwardt, private communications).

\vspace*{-0.5cm}

   \bibliographystyle{klunamed}




\begin{figure}
\centerline{\psfig{file=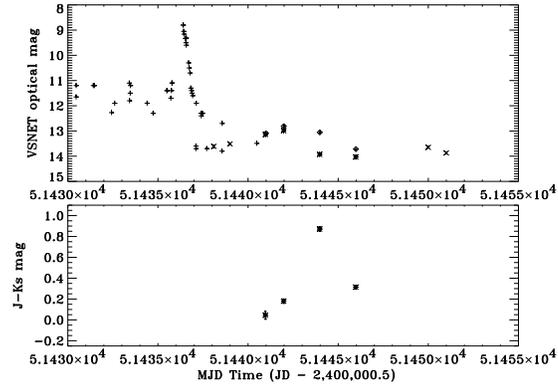,angle=+90.,width=7.3cm}}
\caption[]{Top: +:VSNET, x:V, *:J, diamond:Ks magnitudes). Bottom: J-Ks color.}
\end{figure}

\begin{figure}
\centerline{\psfig{file=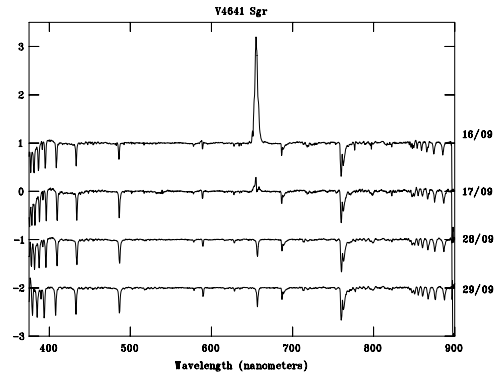,angle=+90.,width=7.3cm}}
\caption[]{Normalized and offset optical spectra.}
\end{figure}

\begin{figure}
\centerline{\psfig{file=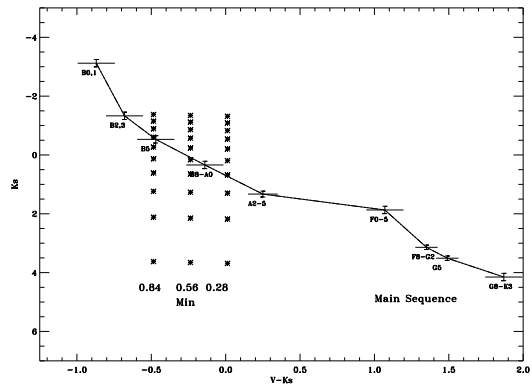,angle=+90.,width=7.3cm}}
\caption[]{Color-magnitude [V-Ks,Ks] diagram. *: Min magnitudes of V4641 Sgr.}
\end{figure}

\end{article}

\end{document}